\documentclass[11pt]{article}
\usepackage[margin=1in]{geometry}
\usepackage{graphicx}
\usepackage{natbib}
\begin{document}
\renewcommand{\thefootnote}{\fnsymbol{footnote}}
\setcounter{footnote}{1}

\title{Space VLBI 2020: Science and Technology Futures\\
  Conference Summary\footnote{\copyright 2020.  All rights reserved}}
\author{%
  T.~Joseph~W.~Lazio ({\small Jet Propulsion Laboratory, California Institute
  of Technology}),\\
  Walter Brisken ({\small National Radio Astronomy Observatory}),\\
  Katherine Bouman ({\small California Institute of Technology}),\\
  Sheperd Doeleman ({\small Harvard-Smithsonian Center for Astrophysics}),\\
  Heino Falcke ({\small Radboud University}), \\
  Satoru Iguchi\\ ({\small Graduate University of Advanced Studies; National Astronomical Observatory of Japan}), \\
  Yuri Y.~Kovalev\\ ({\small Astro Space Center of Lebedev Physical Institute; Moscow Institute of
Physics \& Technology}),\\
  Colin J.~Lonsdale ({\small Massachusetts Institute of Technology/Haystack Observatory}), \\
  Zhiqiang Shen ({\small Shanghai Astronomical Observatory}),\\
  Anton Zensus ({\small Universit\"at zu K\"oln; Max-Planck-Institut fuer Radioastronomie}), \\
  Anthony J.~Beasley ({\small National Radio Astronomy Observatory})}

\maketitle
\renewcommand{\thefootnote}{\arabic{footnote}}

\section{Introduction}\label{sec:intro}

The ``Space VLBI 2020: Science and Technology Futures'' meeting was
the second in The Future of High-Resolution Radio Interferometry in
Space series.  The first meeting
(2018 September~5--6; Noordwijk, the Netherlands) focused on the
full range of science applications possible for very long baseline
interferometry (VLBI) with space-based antennas.
Accordingly, the observing frequencies (wavelengths) considered ranged
from below~1~MHz ($> 300$~m) to above~300~GHz ($< 1$~mm).  For this
second meeting, the focus was narrowed to mission concepts and the
supporting technologies to enable the highest angular resolution
observations at frequencies of~30~GHz and higher ($< 1$~cm).

This narrowing of focus was driven by both scientific and technical
considerations.  First, results from the RadioAstron mission and the
Event Horizon Telescope (EHT) have generated considerable excitement
for studying the inner portions of black hole (BH) accretion disks and
jets and
testing elements of the General Theory of Relativity (GR).  Second,
the technologies and requirements involved in space-based VLBI differ
considerably between~100~MHz and~100~GHz; a related consideration is
that there are a number of existing instruments or mission concepts
for frequencies of approximately 100~MHz and below, while it has been
some time since attention has been devoted to space VLBI at
frequencies above~10~GHz.

This conference summary attempts to capture elements of presentations
and discussions that occurred.\footnote{
\raggedright%
Presentations are available at
\texttt{http://www.cvent.com/events/space-vlbi-2020/agenda-c7b30de0068e454591a66e5639f86668.aspx}.}
While every
effort has been made to summarize material from the formal
presentations, the structure of the meeting included a number of
discussion sessions and a poster session, and not all discussions or
material from those sessions may be included here.
(Appendix~\ref{app:schedule} shows the workshop schedule.)

For reference, standard nomenclature is used for orbits, namely
low-Earth orbit (LEO) having altitudes of no more than about~2000~km
and orbital periods of approximately 90~min.,
medium-Earth orbit (MEO) having altitudes of order 10\,000~km and
orbital periods ranging from a few hours to more than 10~hr, and 
geosynchronous orbit (GEO) having altitudes of approximately 35\,000~km
and orbital periods of just under 24~hr.  Unless otherwise noted, L2
refers to the Earth-Sun Lagrange~2 point, exterior to the Earth's
orbit.

\begin{figure}
  \centering
  \includegraphics[width=0.97\textwidth]{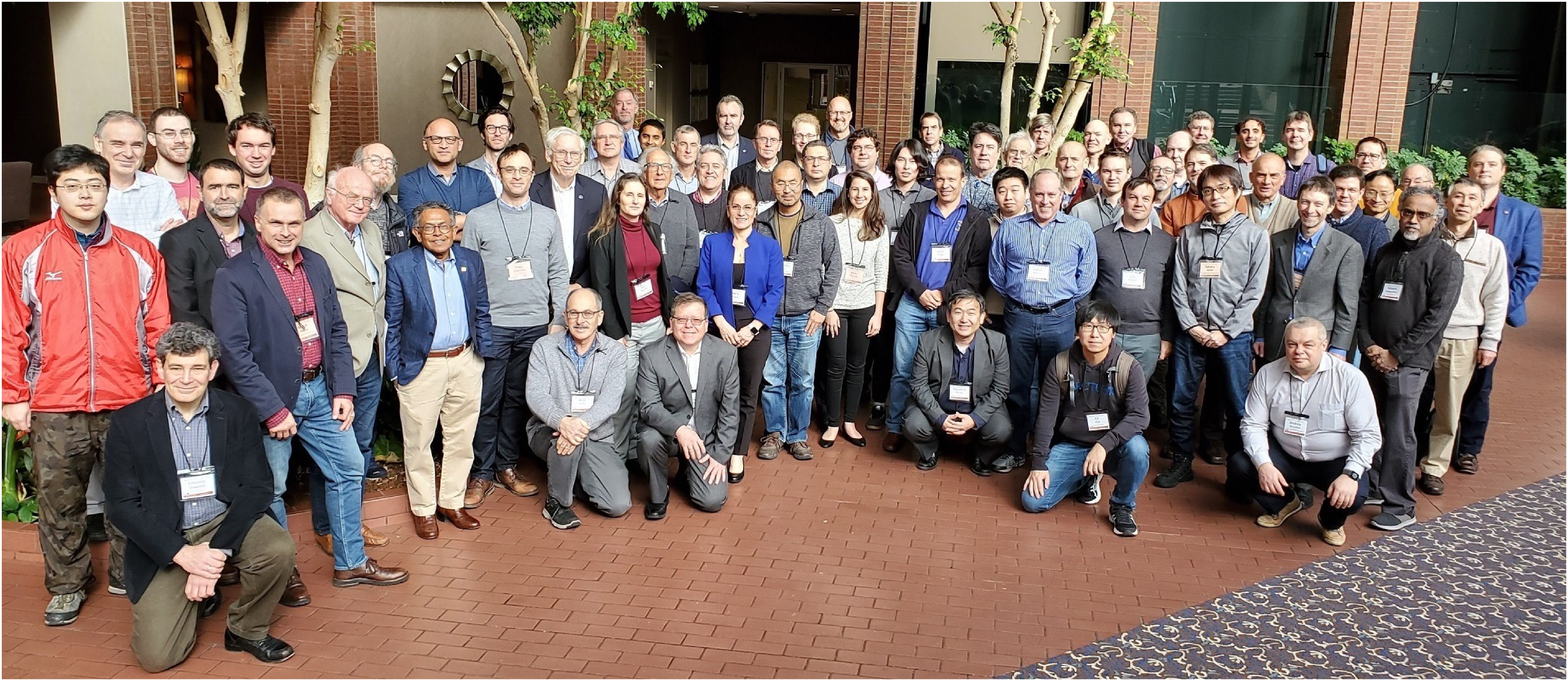}
\end{figure}

\section{Science Motivations}\label{sec:science}

This meeting began with a review of the topics covered at the
first meeting.  Historically, one of the strong drivers for VLBI has
been a ``cosmic conspiracy.''  In order to reach the limiting
brightness temperature for an incoherent synchrotron source
($10^{12}$~K), an interferometric baseline of order the Earth diameter
($\sim 10^4$~km) is required (assuming a source flux density~$S \sim
0.5$~Jy at a redshift $z \sim 1$).  A notable result from RadioAstron,
referenced by multiple speakers, is the detection of a number of
sources with brightness temperatures (well) above the $10^{12}$~K
value expected for an incoherent synchrotron source \citep[e.g.,][]{kkk+16}.
The presence of
such compact sub-components within sources suggests additional physics
is needed, such as coherent emission or highly relativistic beamed
emission.

Subsequently, though, a number of drivers for long baselines,
including extending to space, have emerged.  Broadly, these drivers
have three classes, ``classical'' drivers (e.g., study of jets,
extragalactic masers, astrometry), ``emerging'' drivers (e.g.,
localization of fast radio bursts [FRBs] or electromagnetic
counterparts to gravitational wave events), and ``discovery science''
(of sources yet to be identified).

The focus of this workshop was motivated by recent successes in
imaging a BH's event horizon and the subsequent questions that it has
raised: The role of magnetic fields in launching and collimating
relativistic jets remains uncertain; increasing the sample of targets
beyond Sgr~A* and M87* potentially would allow for population studies;
and it may be possible to test theories of gravity beyond GR by
high-fidelity imaging of a BH's ``photon ring.''

The scientific benefit of being able to measure precisely the diameter
of a BH's ``photon ring'' was discussed by multiple speakers.  The
photon ring marks the region where photons circle the BH one or more
times before escaping to the observer; in actuality, there are a
series of photon rings, progressively thinner and dimmer approaching
the event horizon marking where photons circle once, twice, thrice,
and so on.  For a non-spinning BH of mass~$M$, the photon ring radius
is
\begin{equation}
r_p = \frac{3GM}{c^2} = \frac{3}{2}r_g,
  \label{eqn:ring}
\end{equation}
where $G$ is the Newtonian constant of gravitation, $c$ is the speed
of light, and $r_g$ is the gravitational or Schwarzschild radius.  For
a spinning BH, the intensity of the photon ring becomes highly asymmetrical, raising the
possibility that the BH spin could be measured from the shape of its
photon ring.

Speakers described how photon rings could be used as cosmic rulers and
potentially would enable precision tests of \hbox{GR}.  A potential
confounding aspect of these measurements is that one would need to be
able to distinguish effects due to the astrophysics of the accretion
flow and effects due to the theory of gravity.  Discriminating between
these effects might be difficult because the assumed theory of gravity
could affect the accretion flow in turn affecting the appearance of
the photon ring.

The current EHT results achieved approximately 10\% precision on the
measurement of the M87* photon ring.  Obtaining much higher angular
resolution than did the EHT (e.g., $10\times$ improvement) would
increase dramatically the population of BHs for which photon rings
might be measurable, both by increasing the volume and by being able
to measure much lower masses (e.g., M31*).

A topic noted, though not discussed at length, is how to present
photon ring measurements.  While photon rings have a distinctive
signature in the visibility domain (Fourier or $u$-$v$ plane), neither
the larger astronomical community nor the general public is accustomed
to thinking in the visibility domain.

Jets are ubiquitous structures, observed from pre-main sequence stars
to supermassive BHs (SMBHs).  A continuing question is whether the jets
launched from the immediate environments of BHs are powered by the
Blandford-Znajek or the Blandford-Payne mechanism and what aspect of
the accretion flow or the BH itself determines the mechanism.

Numerous speakers referred to the advances made possible by the high
angular resolution obtained by RadioAstron for probing the inner
structures of jets.  Some RadioAstron observations are consistent with
a frequency-dependent jet structure (limb-brightened at~1.6~GHz,
centrally brightened ``spine'' at~5~GHz).  However, it is also clear
that the innermost structures remain optically thick, even at
frequencies as high as 22~GHz and higher frequency observations are
required to probe closer to the central engine.

Various speakers discussed how to probe GR by imaging the innermost
portion of the accretion flow, near the innermost stable circular
orbit (ISCO).  The time domain behavior of an image could be a
powerful means of determining properties of the inner accretion flow
or the BH or both, particularly if material plunging into the BH can
be identified, from which tests of GR could be conducted.

The importance of fully polarized observations (Stokes~\hbox{I},
\hbox{Q}, \hbox{U}, \hbox{V}) for characterizing the inner portions of
the accretion flow and the magnetic field in the disc was stressed by
multiple speakers.  General Relativistic-magnetohydrodynamic (GRMHD)
simulations are being developed by multiple groups, with radiative
transfer then used to predict what an observer at infinity would
observe.  One potential complication of interpretation is that the
GRMHD simulations are typically only for protons (ions), whereas the
radio emission is generated by electrons.  Differences in temperature
between the electrons and protons, and how to include those
differences in the modeling, likely introduces ambiguities in the predictions.

These various drivers may lead to two different sets of requirements
for any future mission concept.  Some of these science questions would
favor (much) higher angular resolution, requiring longer
interferometric baselines, while others would require higher imaging
fidelity, which could be achieved by exploiting orbital dynamics to
obtain rapid filling of the (Fourier) visibility plane.

There was limited discussion of multi-wavelength and multi-messenger
astronomy opportunities from BH imaging, but one exciting possibility
involves nanohertz gravitational waves, such as being sought by the
North American Nanohertz Observatory for Gravitational Wave Astronomy
(NANOGrav).  A direct example of multi-messenger astronomy would be to
image a SMBH binary that had been identified by NANOGrav as a
gravitational wave emitter.  Alternate possibilities include
discriminating between potential candidate gravitational wave emitters
and providing potential targets for NANOGrav.  More generally, the
formation of a binary SMBH requires some amount of interaction with
its environment to ``harden'' the binary to the point at which
gravitational waves begin to extract sufficient energy from the system
that a future merger is inevitable.  While the binary SMBHs that
NANOGrav detects are too far from merger to be relevant for the
planned Laser Interferometer Space Antenna (LISA) mission, the
detection of binary SMBHs will provide information about merger rates
and potential constraints on the nature of their interactions with
their environments (e.g., stars vs.\ gas).

The International Celestial Reference Frame (ICRF) and the
International Terrestrial Reference Frame (ITRF) are critical
infrastructure for both astronomy and larger society, e.g., global
navigation satellite systems (GNSS) such as the Global Positioning
System (GPS) depend upon precise reference frames in order to obtain
accurate and precise locations on the Earth.  Structures of
astronomical sources used to define reference frames is a recognized
potential systematic, particularly if they change with time.  As such,
merely increasing the number of sources~$N$ used in defining a
reference frame may not improve its precision as $\sqrt{N}$ because
systematic errors would increasingly limit the reference frame. 
Multiple
speakers noted that sub-structure was found by RadioAstron and that a
future space VLBI mission could provide information about source
structure that in turn could be used to improve modeling of sources
used for defining reference frames.

\bigskip 

\begin{figure}[tb]
  \centering
  \includegraphics[width=0.47\textwidth]{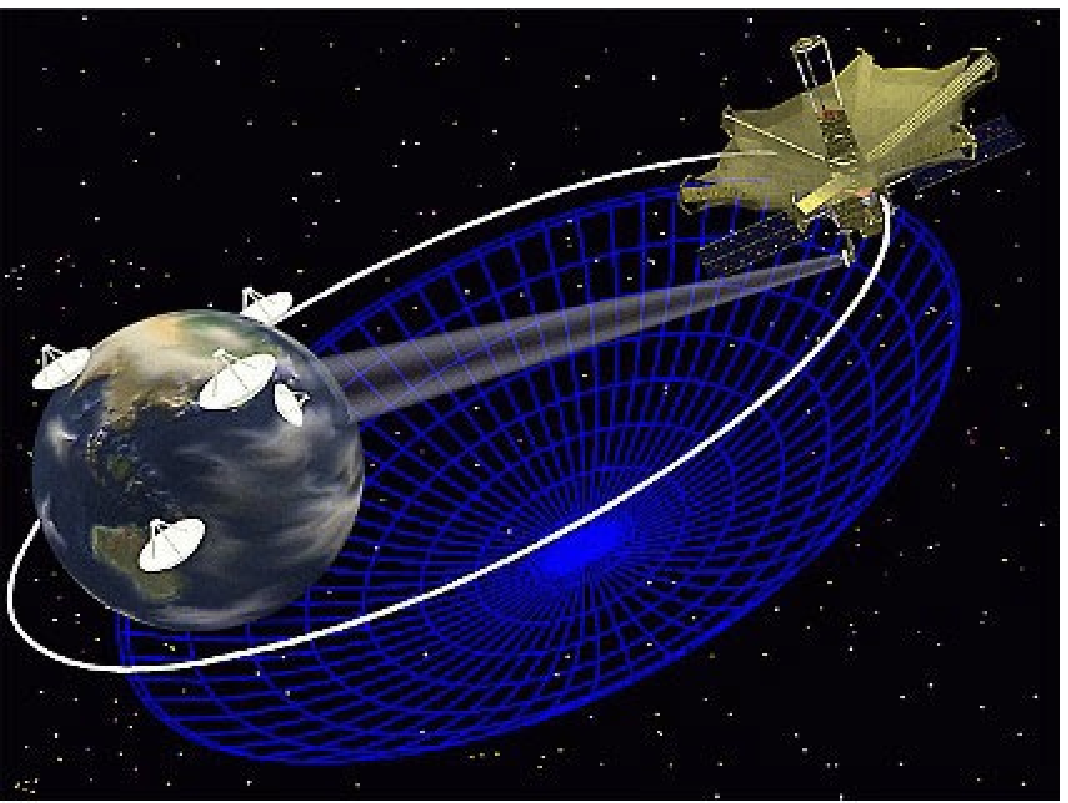}\hfil%
  \includegraphics[width=0.47\textwidth]{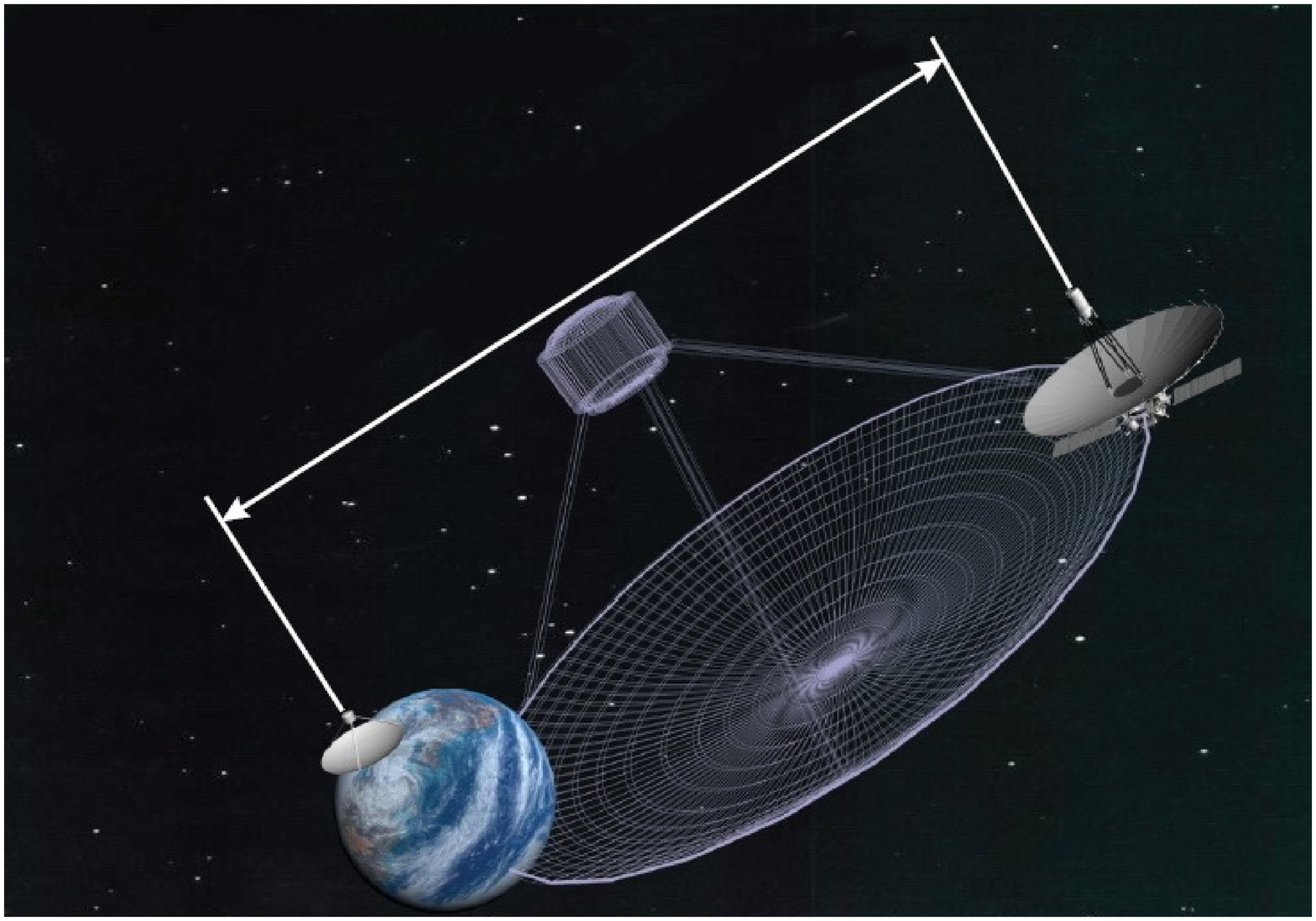}
  \vspace{-1ex}
  \caption{(\textit{left}) VLBI Space Observatory Programme
    (VSOP)/Highly Advanced Laboratory for Communications and Astronomy
    (HALCA) (Courtesy of JAXA); (\textit{right}) RadioAstron (Courtesy
    of Roscosmos).}
  \label{fig:spacevlbi}
\end{figure}

For reference the two prior space VLBI missions, the VLBI Space
Observatory Programme (VSOP)/Highly Advanced Laboratory for
Communications and Astronomy (HALCA) and RadioAstron, were reviewed
(Figure~\ref{fig:spacevlbi}).  The intent was to provide a summary of both the scientific results and
the ``lessons learned'' from these missions.  Not covered during the
meeting, but of importance for the initial demonstrations of the
feasibility of space \hbox{VLBI}, were a series of observations using a
Tracking and Data Relay Satellite System (TDRSS) antenna as the
orbiting component of a VLBI array
\citep{llu+86,lle+89,llu+89,lle+90}.

The VSOP/HALCA mission (1997--2005) was formally an engineering
demonstration mission designed to conduct a number of tests
\citep{vsop}, including the deployment and use of an antenna,
demonstrating pointing accuracy, and the supply of frequency signal
with sufficient precision to the spacecraft.  The spacecraft carried
an 8~m-diameter deployable antenna, on an orbit that took it to an
apogee of~22\,500~km, obtaining angular resolutions of~0.75~mas
at~1.6~GHz and~0.25~mas at~5~GHz.  Initial results from HALCA have
confirmed the presence of multiple AGN having brightness temperatures
in excess of the $10^{12}$~K limit, and re-analysis of the data have
shown that AGN jets can be traced to as close as $300\,r_g$ of the
\hbox{SMBH}.  Further, there are continuing efforts to make the data
more accessible.  One cautionary aspect of the discussion was that,
while HALCA had many notable results, there was concern about the
relatively small number of refereed journal publications that were
produced and that the angular resolution obtained were lower than what
has been obtained in subsequent millimeter-wavelength \hbox{VLBI}.

Following HALCA was to be the ASTRO-G mission, also known as
VSOP2.  Its objectives were to obtain a factor of~10 improvement in
sensitivity, angular resolution, and highest frequency of observation.
It would have consisted of a spacecraft carrying a 10~m-diameter
antenna and cryogenic ally-cooled receivers on an orbit with an apogee
of~30\,000~km.  It would have observed at frequencies of 8~GHz,
22~GHz, and 43~GHz, obtaining resolutions of~260~$\mu$as, 9~$\mu$as,
and~5~$\mu$as, respectively.  However, the level of technology
development required resulted in schedule delays and cost growth, and
the eventual termination of the project.

The RadioAstron mission \citep{radioastron} consisted of a space-borne
10~m diameter antenna equipped with receiving systems at~0.3~GHz,
1.6~GHz, and 22~GHz (as well as a non-operational 5~GHz system).  In
conjunction with ground-based antennas, it obtained angular
resolutions as high as 8~$\mu$as, due to its high apogee
of~350,000~km.  The science mission consisted of observations of
active galactic nuclei (AGN), masers, and pulsars, with multiple
speakers presenting results from different projects.\footnote{%
\raggedright%
A comprehensive list of publications from the RadioAstron mission is
at \texttt{http://www.asc.rssi.ru/radioastron/publications/publ.html}.
}
One of its prime objectives was to test for the presence of AGN having
brightness temperatures in excess of the incoherent synchrotron limit,
as discussed above.  The distribution of AGN brightness temperatures
does extend above this limit, with some sources having brightness
temperatures up to $10^{14}$~\hbox{K}.

Among other results from RadioAstron are that linear polarization
observations could be conducted when the satellite was close to
periastron.  As noted above, polarization measurements are essential
to characterizing the magnetic field, and the detection of and
subsequent study of polarized emission on the sub-milliarcsecond
scales was an expected result from RadioAstron.  Observations of
pulsars revealed complex sub-structure in images, consistent with
interstellar scattering along the line of sight.  Finally, with the
deployment of km-scale neutrino detectors, it was discovered that AGN
that are particularly bright to RadioAstron are also potential
neutrino sources.

\section{Technology Drivers and Readiness}\label{sec:tech}

The meeting surveyed the state of technology development and readiness
for future space VLBI missions.  This survey was wide ranging, from
the receiving antenna to the data reception on the ground, and a
conclusion of the meeting was that advancing some technologies would
be of general use to any future mission while the advancement of other
technologies would only be relevant for a limited set of mission
concepts (Table~\ref{tab:techdevelop}).

\begin{table}[tbh]
  \centering
  \caption{Space VLBI Technology Classification\label{tab:techdevelop}}
  \begin{tabular}{ll}
   \noalign{\hrule\hrule}
   \textbf{Likely Mission Agnostic} & \textbf{Likely Mission Specific} \\
   \noalign{\hrule}
   Sampling \& Digital Processing & Antenna (size, frequency) \\
   Data Transmission              & Receiver (bands, cooling) \\
   Time \& Frequency References   & Orbit Design \& Orbit
                                    Determination (precision) \\
                                  & Data Analysis (imaging, modeling,
                                    polarization) \\
   \noalign{\hrule\hrule}
  \end{tabular}
\end{table}

Various speakers noted that the likely time scale for the development
and qualification of a new component is plausibly a decade.  However,
a tentative consensus from the workshop is that, while there are
likely to be challenges with some aspects of technology, many
technologies needed for at least some kinds of space VLBI missions are
reasonably mature or will be matured in various technology
demonstration missions over the next few years.  (``No unobtainium
appears required.'')

The following provides a brief summary of various points discussed.
It is intended to be illustrative of the range of topics, but there
are undoubtedly other aspects of potential mission concepts that were
not discussed at length.  Most notably, there was little discussion of
what ground data processing (including, but perhaps not limited to
imaging) would be required.
\begin{description}
\item[Antenna]%
  Both HALCA and RadioAstron used deployable antennas, and Millimetron
  (\S\ref{sec:mission}) is planning to do so, while sub-millimeter
  wavelength missions have used monolithic antennas.  None of the
  space VLBI deployable antennas have been required to operate
  above~25~GHz.  Antennas with diameters up to~20~m and highest
  operating frequencies up to~30~GHz are feasible.  Some capabilities
  for deployable antennas up to~50~GHz have been demonstrated, and
  there is initial testing of antennas operating up to~110~GHz.

\item[Receivers]%
  Presentations and posters described technologies for receivers.  New
  low-noise amplifier (LNA) technology, based on kinetic inductance
  detectors (KIDs), are being developed for millimeter-wave
  paramplifiers, for both ground- and space applications.  New
  receiving systems are being developed and deployed to telescopes
  around the world, telescopes that could potentially serve as members
  of a ground telescope network to complement space-based antennas.
  These telescopes included the Atacama Pathfinder Experiment (APEX)
  and those in the Korean VLBI Network (KVN), among others.

\item[Frequency Standards and Clocks]%
  Frequency stability and time transfer are critical elements of any
  VLBI array in order to ensure accurate correlation.  For mission
  concepts involving a combination of space- and ground-based
  antennas, the space-based antennas will be moving at (much) higher
  velocities relative to the ground-based antennas and potentially
  through a varying gravitational potential.  Maintaining the
  environmental stability (e.g., temperature) for a space-based clock
  is likely to be more resource demanding relative to a ground-based
  clock.  Further, the requirements for VLBI may lead to requirements
  on space clock stability on time scales that are shorter than many
  other applications.  The RadioAstron mission demonstrated the use of
  a space-based hydrogen maser, and there are a number of new clock
  developments occurring, such as the Deep Space Atomic Clock (DSAC),
  and developments in sub-nanosecond time transfer accuracy that
  will likely be promising for future space \hbox{VLBI}.

\item[Orbit Determination]%
  Just as for ground-based VLBI arrays, space-based antenna positions
  must be constrained prior to correlation so as to avoid excessive
  searching in phase delay or rate.  For a space-based antenna,
  position determination becomes a problem in orbit determination
  because of the (high) orbital velocities involved.  For instance, an
  antenna in LEO (velocity $\approx 7$~km~s${}^{-1}$) travels a
  distance of~1~mm (equivalent observing frequency $= 300$~GHz)
  in~140~ns.

  State-of-the-art orbit determination for single satellites located
  below GNSS constellations is at the centimeter level for absolute
  measurements and at the millimeter level for relative measurements
  between two spacecraft.  The highest (relative) precision obtained
  is likely to be between the two GRACE-FO spacecraft (relative
  separation $\approx 200$~km), which can be of order 1~$\mu$m.  There
  is considerably less experience with spacecraft above the GNSS
  constellations, with the possible exception of the lunar-orbiting
  dual spacecraft Gravity Recovery And Interior Laboratory (GRAIL)
  mission.  Judgments by workshop participants suggest that orbit
  determination above the GNSS constellations is likely to be at least
  an order of magnitude worse, at least in the near term, as compared
  to that below GNSS constellations.

\item[Data Transfer]%
  Data rates for ground-based VLBI are routinely in the 2~Gbps regime
  and the EHT has achieved rates of 64~Gbps.  Ground-based VLBI data
  rates are being achieved currently, by the European Data Relay
  Service (1.8~Gbps) and, soon, by NASA-India Space Research
  Organization (ISRO) Synthetic Aperture Radar (NISAR) mission (up
  to~4~Gbps).  Over the next few years, technical demonstration
  missions are planned that should achieve as high as 100~Gbps from
  \hbox{GEO}, typically by using laser (optical) communications.  All
  other things being equal, data rates scale as $R^{-2}$ for a
  range~$R$.
\end{description}

\section{Mission Concepts}\label{sec:mission}

The Millimetron mission concept involves a 10~m deployable and cooled
sub-mm and far-IR telescope.  Placed into an orbit around the
Earth-Sun L2 point, it will conduct VLBI observations at wavelengths
of~0.8~mm to~3~mm (350~GHz to~100~GHz) and single dish observations to
wavelengths as short as 80~$\mu$m.  The antenna will be cooled for the
first three years of the expected 10~year science operations duration
of the mission.  The concept is in a Phase~A study, funded by the
Russian Space Agency.

The deployable antenna is made of a series of panels.  The panel
production process has been qualified, with a substantial reduction in
areal mass density having been achieved relative to many previous and
current sub-mm and far-IR missions,
and panel production is beginning.  The antenna itself is considered
to be in Phase~\hbox{B}.  Development of instruments is under
discussion, including with international partners, but space VLBI
receivers are planned to be part of the instrument package.  The
planned recording rate on-board the spacecraft is 2~Gbps for four
channels, for a total rate up to 16~Gbps, but the downlink data rate
is only 1.6~Gbps.

\bigskip

For future missions, two general architectures were discussed.  In one
architecture, one or more orbiting antennas have relatively low orbits
(from LEO to MEO).  The objective of such an architecture is to obtain
higher imaging fidelity, potentially on more rapid time scales, in
order to track dynamic changes in the inner accretion disk and jet.
This approach would be most beneficial for lower-mass BHs (e.g.,
Sgr~A*), for which time scales are typically minutes to days.  If
there are two or more orbiting antennas, it may be possible to realize
an entirely space-based array, with no dependency on any ground-based
telescopes.  Such a space-based array could also operate at
frequencies at which the atmosphere is largely or essentially opaque
($> 600$~GHz).

In the second mission architecture, a single antenna is at a large
distance from the Earth (e.g., Earth-Moon L2 or Earth-Sun L2).  The
objective of such an architecture is to obtain extreme angular
resolutions, in combination with ground-based telescopes, in order to
measure the characteristics of the ``photon rings'' for a large number
of SMBHs.
Some estimates for the number of SMBHs for which the ``photon ring''
characteristics could be measured were as high as $10^4$ objects.

There was also discussion of a mixed architecture in which an orbiting
antenna was placed on a highly eccentric orbit.  During most of its
orbit, it would be at a large distance from the Earth so as to obtain
an extreme angular resolution while, for a short portion of the orbit,
its rapid orbital motion covers the (Fourier) visibility plane rapidly
in order to improve image fidelity.

Both HALCA and RadioAstron had significant international partnerships.
The U.{}S.\ Space VLBI Project supported U.{}S.\ community engagement
with both HALCA and RadioAstron, though RadioAstron launched after the
termination of the project.  Key ``lessons learned'' from the U.{}S.\
Space VLBI Project include
\begin{itemize}
\item A project can have non-scientific motivations.  
  The U.{}S.\ Space VLBI Project was motivated partially by
  high-level U.{}S.\ Government desire to improve relations between
  the U.{}S.\ and the then Soviet Union.
\item At least in the U.{}S., prior to a formal project, there is
  often a pre-project phase, and this series of meetings could be
  considered an example of a pre-project activity.
\item The Project had a memorandum series, some of the analysis reported
  in those memoranda may be of value in developing future missions,
  and there was interest expressed in having those memoranda being
  available publicly.
\item Over-optimism about technology maturity can lead
to cost overruns, which can lead to severe stress or even mission
cancellation.
\item Debates about whether resolution or image quality are the most
  important factor can be acrimonious.
\item It is important to have and adhere to a clear message regarding
  the science and technology in order to obtain high-level support,
  from both the science community and management (at institutes and
  within space agencies).
\item It is important to discuss the distribution of science benefits
  among mission partners.
\item The existence of an international science council (e.g., the
  RadioAstron International Science Council [RISC]) is important in
  obtaining long-term and broad engagement.
\item Establishing an early agreement by ground-based radio
  telescopes, if they are to enable the mission's science, is
  critical.  There was some discussion that, prior to the
  establishment of the \hbox{VLBA}, the science case for space VLBI
  may have been perceived as weak.
\item Next-generation missions need to have fewer constraints imposed
  by the spacecraft or operations in order to obtain better science.
\end{itemize}

Two ground-based telescopes of relevance for future space-based VLBI
are the Atacama Large Millimeter/submillimeter Array (ALMA) and the
next-generation Very Large Array (ngVLA).  Due to the addition of
phasing capabilities \citep{mcd+18}, ALMA was already an integral part
of the initial EHT results on M87*, and it is now part of the Global
mm VLBI Network (GMVA) and the \hbox{EHT}.  ALMA has developed a
development roadmap to~2030 \citep{alma2030}, and an on-going project
to improve its capability involves improving the capabilities to
include (i)~Spectral line \hbox{VLBI}; (ii)~Extending the frequency
range of the current phasing mode to Bands~1--7; (iii)~Improving the
calibration to enable observations of weaker sources;
and~(iv)~Single-dish mode and pulsar modes.

The ngVLA is a concept for a North American-based telescope operating
between approximately 1~GHz and~120~GHz \citep{ngvla}, and it would be
a clear complement to any future space-based VLBI system operating
below~100~GHz.  Its design includes antennas across North America for
trans-continental VLBI and a phasing mode for pulsar observations.
\section{Future Steps}\label{sec:future}

Any future mission will require a compelling science case, and
continued millimeter-wave VLBI on the ground will be an essential
aspect of building that science case.  Notably future EHT
observations, particularly of Sgr~A*, could help motivate whether a
mission to achieve high image fidelity or extreme angular resolution
is more compelling.

A number of technology advances are likely to occur in the near term,
with motivations separate from space \hbox{VLBI}, but from which a
space VLBI mission would benefit.  The DSAC technical demonstration
mission was noted above, but there also are a number of technical
demonstrations of laser (optical) communications to achieve higher
data rates.  Monitoring of these technical developments, potentially
complemented by proposals for targeted technical developments in other
areas relevant to a space VLBI mission, is warranted.  There was also
substantial discussion of potential pathfinder missions, in order to
prove out technologies, and discussions about the extent to which
(long-duration) balloons would be able to contribute either to the
technology development or even science observations.

At various times during the meeting, the extent to which the
participants adequately represented the diversity of the astronomical
community was questioned.  In some potential axes of diversity, there
was broad representation, but other axes were quite narrow.

Finally, there was general agreement that a third meeting in the
series should occur, likely in the latter half of~2021.\footnote{
This conclusion is likely to have to be revisited in light of the
novel coronovirus pandemic.}
On that time
scale, a number of events should occur including future maturation of
the Millimetron concept, new results from the EHT including on Sgr~A*,
and the U.{}S.\ Decadal Survey report.  With those developments, it
may be possible to identify a concrete next step for a future space
VLBI mission.

\paragraph{Acknowledgments}\label{par:acknowledge}

We thank K.~Prairie for her expert organization of the conference.
This work made use of NASA's Astrophysics Data System Bibliographic Services.
The National Radio Astronomy Observatory is a facility of the National Science Foundation operated under cooperative agreement by Associated Universities, Inc.
The RadioAstron project was led by the Astro Space Center of the Lebedev Physical Institute of the Russian Academy of Sciences and the Lavochkin Scientific and Production Association under a contract with the State Space Corporation \hbox{ROSCOSMOS}, in collaboration with partner organizations in Russia and other countries.
Part of this research was carried out at the Jet Propulsion
Laboratory, California Institute of Technology, under a contract with
the National Aeronautics and Space Administration.

\clearpage

\appendix

\section{Workshop Schedule}\label{app:schedule}

\begin{center}
\includegraphics[width=0.9\textwidth]{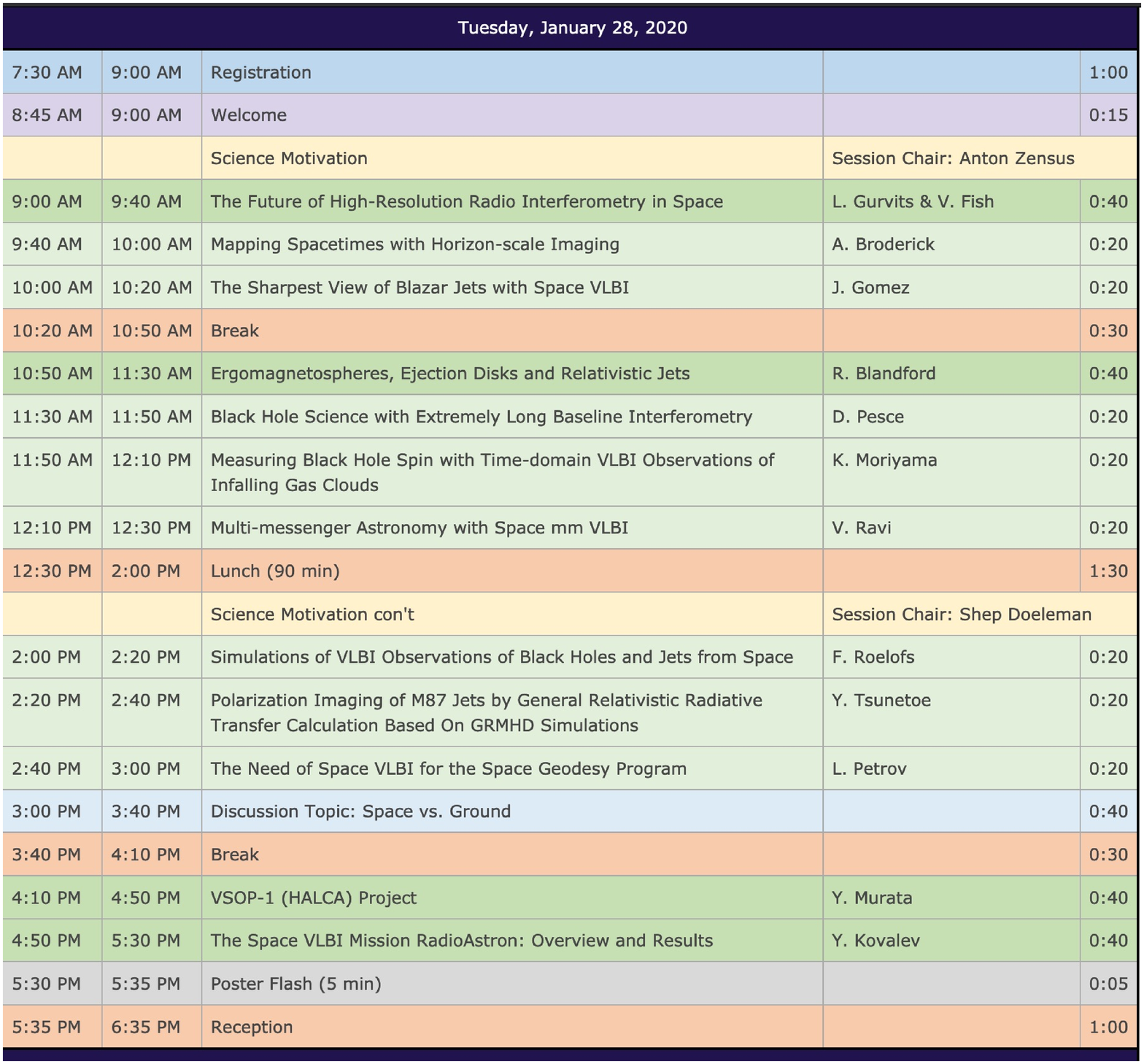}
\end{center}
\clearpage

\begin{center}
\includegraphics[width=0.9\textwidth]{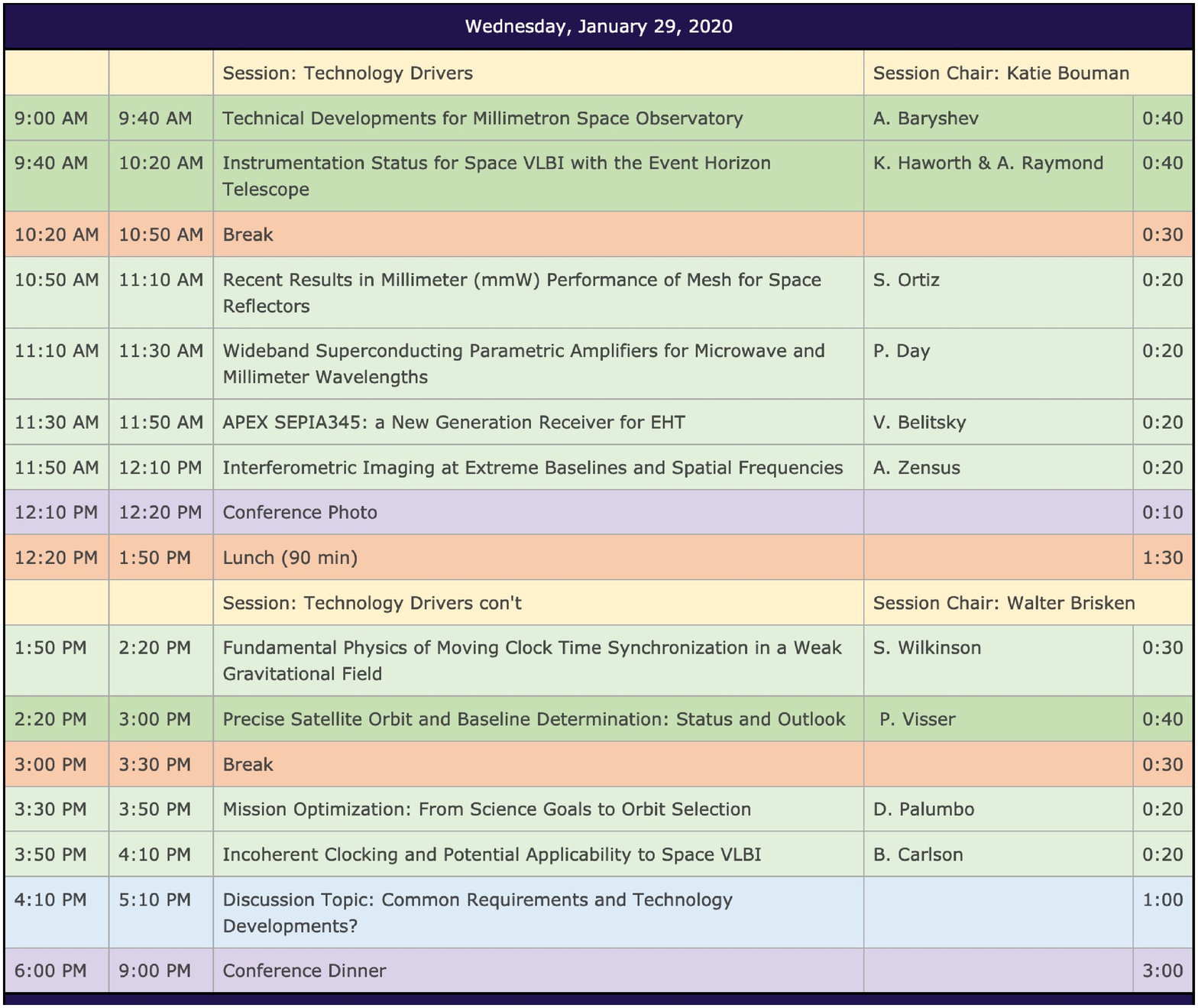}
\end{center}

\clearpage

\begin{center}
\includegraphics[width=0.9\textwidth]{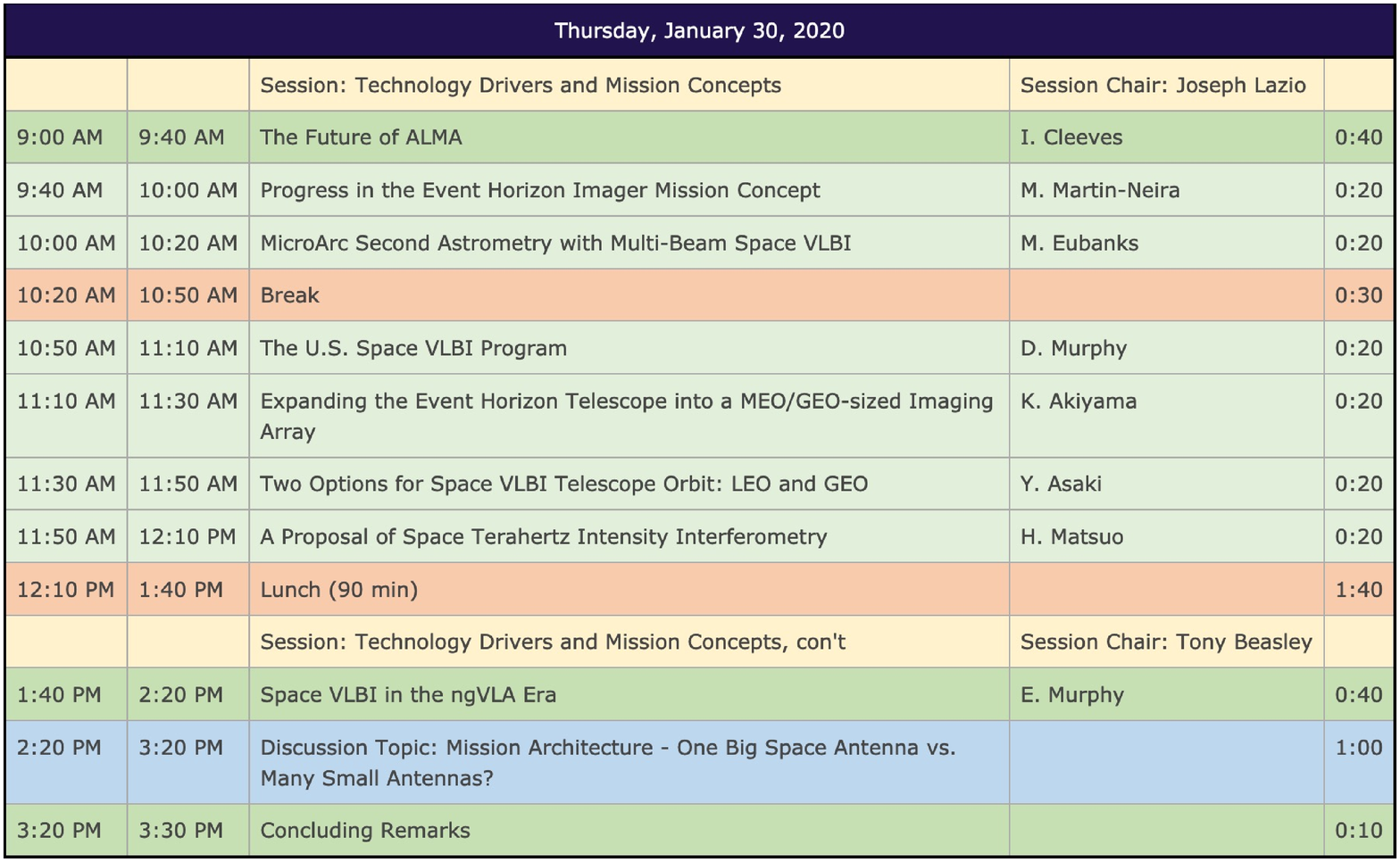}  
\end{center}

\end{document}